\documentclass[12pt]{article}

\begin{document}

\def\CA{{\cal A}}
\def\CB{{\cal B}}
\def\CD{{\cal D}}
\def\CE{{\cal E}}
\def\CF{{\cal F}}
\def\CG{{\cal G}}
\def\CH{{\cal H}}
\def\CI{{\cal I}}
\def\CJ{{\cal J}}
\def\CK{{\cal K}}
\def\CL{{\cal L}}
\def\CM{{\cal M}}
\def\CN{{\cal N}}
\def\CO{{\cal O}}
\def\CP{{\cal P}}
\def\CQ{{\cal Q}}
\def\CR{{\cal R}}
\def\CS{{\cal S}}
\def\CT{{\cal T}}
\def\CU{{\cal U}}
\def\CV{{\cal V}}
\def\CW{{\cal W}}
\def\CX{{\cal X}}
\def\CY{{\cal Y}}
\def\CZ{{\cal Z}}

\newcommand{\todo}[1]{{\em \small {#1}}\marginpar{$\Longleftarrow$}}
\newcommand{\labell}[1]{\label{#1}\qquad_{#1}} 
\newcommand{\bbibitem}[1]{\bibitem{#1}\marginpar{#1}}
\newcommand{\llabel}[1]{\label{#1}\marginpar{#1}}

\newcommand{\sphere}[0]{{\rm S}^3}
\newcommand{\su}[0]{{\rm SU(2)}}
\newcommand{\so}[0]{{\rm SO(4)}}
\newcommand{\bK}[0]{{\bf K}}
\newcommand{\bL}[0]{{\bf L}}
\newcommand{\bR}[0]{{\bf R}}
\newcommand{\tK}[0]{\tilde{K}}
\newcommand{\tL}[0]{\bar{L}}
\newcommand{\tR}[0]{\tilde{R}}

\newcommand{\btzm}[0]{BTZ$_{\rm M}$}
\newcommand{\ads}[1]{{\rm AdS}_{#1}}
\newcommand{\ds}[1]{{\rm dS}_{#1}}
\newcommand{\eds}[1]{{\rm EdS}_{#1}}
\newcommand{\sph}[1]{{\rm S}^{#1}}
\newcommand{\gn}[0]{G_N}
\newcommand{\SL}[0]{{\rm SL}(2,R)}
\newcommand{\cosm}[0]{R}
\newcommand{\hdim}[0]{\bar{h}}
\newcommand{\bw}[0]{\bar{w}}
\newcommand{\bz}[0]{\bar{z}}
\newcommand{\beq}{\begin{equation}}
\newcommand{\eeq}{\end{equation}}
\newcommand{\beqn}{\begin{eqnarray}}
\newcommand{\eeqn}{\end{eqnarray}}
\newcommand{\pat}{\partial}
\newcommand{\lp}{\lambda_+}
\newcommand{\bx}{ {\bf x}}
\newcommand{\bk}{{\bf k}}
\newcommand{\bb}{{\bf b}}
\newcommand{\BB}{{\bf B}}
\newcommand{\tp}{\tilde{\phi}}
\hyphenation{Min-kow-ski}

\newcommand{\beql}[1]{\begin{equation}\label{eq:#1}}
\newcommand{\eq}[1]{(\ref{eq:#1})}
\newcommand{\ack}[1]{{\bf Pfft!: #1}}\newcommand{\talkpt}[1]{{\bf TP: #1}}
\newcommand{\field}[1]{\mathbb{#1}}
\newcommand{\CC}{{\field C}}
\newcommand{\RR}{{\field R}}
\newcommand{\ZZ}{{\field Z}}

\newcommand{\scale}{\sigma}

\def\apr{\alpha'}
\def\str{{str}}
\def\lstr{\ell_\str}
\def\gstr{g_\str}
\def\Mstr{M_\str}
\def\lpl{\ell_{pl}}
\def\Mpl{M_{pl}}
\def\varep{\varepsilon}
\def\del{\nabla}
\def\grad{\nabla}
\def\tr{\hbox{tr}}
\def\perp{\bot}
\def\half{\frac{1}{2}}
\def\p{\partial}
\def\perp{\bot}
\def\eps{\epsilon}

\def\NPB{{\it Nucl. Phys. }{\bf B}}
\def\PL{{\it Phys. Lett. }}
\def\PRL{{\it Phys. Rev. Lett. }}
\def\PRD{{\it Phys. Rev. }{\bf D}}
\def\CQG{{\it Class. Quantum Grav. }}
\def\JMP{{\it J. Math. Phys. }}
\def\SJNP{{\it Sov. J. Nucl. Phys. }}
\def\SPJ{{\it Sov. Phys. J. }}
\def\JETPL{{\it JETP Lett. }}
\def\TMP{{\it Theor. Math. Phys. }}
\def\IJMPA{{\it Int. J. Mod. Phys. }{\bf A}}
\def\MPL{{\it Mod. Phys. Lett. }}
\def\CMP{{\it Commun. Math. Phys. }}
\def\AP{{\it Ann. Phys. }}
\def\PR{{\it Phys. Rep. }}

\renewcommand{\thepage}{\arabic{page}}
\setcounter{page}{1}

\rightline{hep-th/0212057}
\rightline{VPI-IPPAP-02-16}
\rightline{CERN-TH/2002-351}
\vskip 1cm
\centerline{\Large \bf The Cosmological Constant}\vskip0.25cm
\centerline{\Large \bf and the Deconstruction of Gravity}
\vskip 1cm
\vskip 1cm

\renewcommand{\thefootnote}{\fnsymbol{footnote}}
\centerline{{\bf Vishnu Jejjala,${}^{1}$\footnote{vishnu@vt.edu}
Robert G. Leigh,${}^{2,3}$\footnote{rgleigh@uiuc.edu} and
Djordje Minic${}^{1}$\footnote{dminic@vt.edu}
}}
\vskip .5cm
\centerline{${}^1$\it Institute for Particle Physics and Astrophysics}
\centerline{\it Department of Physics, Virginia Tech}
\centerline{\it Blacksburg, VA 24061, U.S.A.}
\vskip .5cm
\centerline{${}^2$\it CERN-Theory Division}
\centerline{\it CH-1211, Geneva 23, Switzerland}
\vskip .5cm
\centerline{${}^3$\it Department of Physics}
\centerline{\it University of Illinois at Urbana-Champaign}
\centerline{\it 1110 W. Green Street, Urbana, IL 61801, U.S.A.}

\setcounter{footnote}{0}
\renewcommand{\thefootnote}{\arabic{footnote}}

\begin{abstract}
Witten has presented an argument for the vanishing of the cosmological
constant in $2+1$ dimensions.  This argument is crucially tied to
the specific properties of $(2+1)$-dimensional gravity.  We argue that
this reasoning can be deconstructed to $3+1$ dimensions under certain
conditions. Our observation is also tied to a possibility that there
exists a well-defined UV completion of $(3+1)$-dimensional gravity.

\end{abstract}
\newpage

\section{Introduction and Summary}

The cosmological constant has been an enigma in theoretical physics since
it was first realized that in any simple field theoretic notion of quantum
gravity, power divergences lead to large renormalization, sensitive to the
largest scales available in the theory \cite{cc}. In terms of naive 
power-counting, the vacuum energy corresponds to a relevant operator. One 
might go further to
say that a lack of understanding of power divergences is at the root of
each of the basic theoretical problems in particle physics including the
various hierarchy problems and the aforementioned cosmological constant
problem. It is important to realize that in the case of power divergences,
it is not enough to come up with a mechanism for canceling the parameter
at a given scale; rather, it must be canceled all the way into the infrared
(IR).

The cosmological constant problem has become perhaps
even more acute given the recent
astronomical data suggesting the existence of a
positive but small vacuum energy density, being roughly in proportion to the
present mass density of the universe \cite{astro}.  Thus we are faced with
a two-fold cosmological constant problem \cite{cc}: first, why is the vacuum
energy small and second, why is the vacuum energy in proportion to the current
mass density?  In this article we will address the first question.

Field theories that are non-renormalizable (and hence ill-defined in the
ultraviolet (UV)) may be defined through a certain process of dimensional
reduction referred to as deconstruction \cite{nima}. This has been
demonstrated  in theories with internal gauge symmetries, for example, in the 
context of five-dimensional Yang-Mills
theories. One dimension of the IR theory is put on a lattice
and the resulting theory may be thought of  as a Goldstone realization
of a UV four-dimensional gauge theory. 
In this way, the continuum higher dimensional theory is thought of as the {\it 
infrared} limit of a lower dimensional theory. An important aspect of these 
constructions then is motivating why the theory has this infrared behavior.

It is enticing to think of gravity in this context: from a four-dimensional 
field theoretic point of
view, Einstein's general relativity is famously 
perturbatively non-renormalizable. In order to extend the deconstruction 
ideas to gravity, we must confront the spacetime general coordinate and 
Lorentz symmetries.
Thus in this paper, we explore the idea that four-dimensional quantum gravity
may be {\it defined} through deconstruction.  It is particularly convenient to 
work in the vierbein formalism.  It was shown  \cite{wittencs} long ago that 
three-dimensional gravity is a Chern-Simons (CS) gauge theory and thus is a 
well-defined quantum
theory. In particular its UV character is sensible because it
is topological. The deconstruction to $3+1$ dimensions would follow the path 
of regarding a three-dimensional theory (a close cousin of CS gravity coupled 
to matter) as a lattice version of a
four-dimensional theory.

Of course, there exists a rather large body of evidence that gravitational 
theories should be thought of, in some way, as local theories in one fewer 
dimension.
First, a purely gravitational theory has no local degrees of freedom
in the usual sense of a local quantum field theory. In the work of 't
Hooft \cite{thooft} and Susskind \cite{sussholo}, it was realized that
an interpretation of this is that a gravitational theory is holographic
--- the observables are not extensive, but related to co-dimension one
structures. This is of course supported by the thermodynamics of black
holes, where entropy is proportional to the area of the event horizon
\cite{bhent}. But most impressively, the idea has been given a concrete
realization in the $AdS$/CFT construction and its relatives \cite{adscft}.

In the case of four-dimensional gravity, we might then try to follow this
path directly and construct its physics in terms of a three-dimensional
field theory. In realistic models with positive cosmological constant,
this might mean some version of a de Sitter/Euclidean CFT correspondence
\cite{stromds}. Although the existence of such a correspondence has not been
established conclusively, a variety of consequences have been considered
in Refs. \cite{lls}.

Note that the existing holographic duals of gravity,
provided by the $AdS$/CFT correspondence, are defined in terms
of non-gravitational theories.  On the other hand, there are proposals
for a non-perturbative definition of gravity which involve gravitational
degrees of freedom. Perhaps the most notable example of this type is Matrix
theory \cite{matrix}, which in some sense can be viewed as an example of
``bulk'' holography.

Finally, more than twenty years ago Weinberg suggested the idea of
``asymptotic safety'' which essentially advocates the existence of a UV
fixed point for $(3+1)$-dimensional gravity \cite{uvfp}, in the sense of a
Wilson-Fisher $\epsilon$-expansion.\footnote{In particular, by considering
$(2 + \epsilon)$-dimensional gravity.} 
One of the main points of this article is precisely
the suggestion that $(3+1)$-dimensional gravity may indeed have
a short distance fixed point given in terms of $(2+1)$-dimensional gravity coupled
to $(2+1)$-dimensional matter. 

One might be initially puzzled by a suggestion that 
$(3+1)$-dimensional gravity can be defined in terms
of $(2+1)$-dimensional gravity coupled to $(2+1)$-dimensional
matter. After all,
$(3+1)$-dimensional gravity has propagating degrees of freedom.  However,
$(2+1)$-dimensional gravity, viewed as a CS gauge theory, is purely topological.
There are no propagating, local gravitational degrees of freedom.  
How can then
a $(2+1)$-dimensional theory of matter coupled to gravity
account for the local, propagating, $(3+1)$-dimensional
degrees of freedom, such as gravitational waves?
What our proposal suggests is that ``most''
of the degrees of freedom of $(3+1)$-dimensional gravitational theory arise from the non-gravitational part
of its $(2+1)$-dimensional UV completion.
The UV completion of $(3+1)$-dimensional gravity is 
``holographic'' in this sense.

If four-dimensional gravity may be thought of as a three-dimensional
theory in a useful way, what of the cosmological constant? Several
years ago, Witten \cite{wittencc} observed that peculiar properties of
$(2+1)$-dimensional gravity can lead to vanishing vacuum energy in $2+1$
dimensions. No precise mechanism for connecting this to four dimensions has
been presented, although Witten's context was firmly rooted in the duality
between M-theory and the strong coupling limit of Type IIA or heterotic
strings \cite{wittencc2}. Can this mechanism be used instead in our context to provide
insight into the vacuum energy in four dimensions?  In this note we argue
that Witten's reasoning can be deconstructed to $3+1$ dimensions under
very specific conditions.\footnote{This would perhaps imply alternative
interpretations of the recent astronomical data \cite{astro}. Such
interpretations are explored in Ref.\ \cite{nemanja}.}

The crucial observation we make in this paper is that provided one can
define a UV completion of $(3+1)$-dimensional gravity in terms of purely
$(2+1)$-dimensional gravitational and matter data, then the argument of
Witten can be deconstructed to $3+1$ dimensions.  We motivate our argument
by recalling a remarkable fact from {\it classical} general relativity which
states that in the presence of a space-like Killing field, $3+1$ vacuum
general relativity is equivalent to $(2+1)$-dimensional general relativity
coupled to an $SO(2,1)$ non-linear $\sigma$-model \cite{grbook,aashtekar}.  We then
proceed to provide a {\it quantum} analogue of this classical theorem and
argue that a full quantum theory of $(3+1)$-dimensional general relativity
can be defined at short distance in terms of $(2+1)$-dimensional gravity
coupled to $(2+1)$-dimensional matter.  This then provides support for the
claim that Witten's observation about the vanishing $(2+1)$-dimensional
vacuum energy may be also valid in the world we observe.

\newcommand{\pa}{\partial}
\newcommand{\Dslash}[1]{\not\!\!{#1}}
\newcommand{\dslash}[1]{\not\!{#1}}

\section{The Cosmological Constant in $2+1$ Dimensions}

It was observed by Witten \cite{wittencc} that supersymmetry in $2+1$
dimensions can lead to vanishing vacuum energy in the absence of a mass
degenerate spectrum of bosonic and fermionic states.  The vacuum state is
supersymmetric, and therefore the cosmological constant is zero, but the
excited states are not mass degenerate because unbroken global supercharges
do not exist in $2+1$ dimensions \cite{Hen}.  Having unbroken global
supercharges in the theory, which is what leads to the mass degeneracy of
the bose-fermi spectrum in the first place, necessitates the existence
of spinor fields that are covariantly constant at infinity.  In $2+1$
dimensions any excited state gives a conical geometry whose deficit angle
prohibits spinor fields with covariantly constant asymptotics.  Thus,
there is no mass degeneracy of bose-fermi excitations.  The non-degeneracy
of the spectrum of low-energy excitations scales as the inverse power
of the three-dimensional Newton constant under the assumption of weak
gravitational coupling \cite{wittencc2}.

Although a precise realization of Witten's argument about a supersymmetric
vacuum with non-supersymmetric excitations apparently does not exist in the 
literature,\footnote{Witten's argument that there can be a supersymmetric vaccum with
non-supersymmetric excitations has not been lifted to four-dimensions.
Most $(3+1)$-dimensional asymptopia, however, are not consistent with the
existence of globally conserved supercharges.  For example, time-dependent
backgrounds usually do not allow covariantly constant spinors.}
Becker, Becker, and Strominger \cite{bbs} provide an instructive construction
with a solitonic ground state.\footnote{For a related discussion see Ref.\
\cite{fk}.}  We briefly review their considerations.

Becker, Becker and Strominger 
considered an $N=2$ abelian Higgs model in $2+1$ dimensions
\cite{misc} and studied a 
Nielsen-Olesen vortex \cite{no} configuration in
this theory. 
The solitonic configuration breaks half the supersymmetry.  When this model is coupled to
supergravity, the $(2+1)$-dimensional
gravitational background of this
soliton has a particular asymptotic behavior
describing a conical geometry
\begin{equation}
ds^2 = -dt^2 + \frac{dzd\bar z}{|z|^{2M/M_{Pl}}},
\end{equation}
where $M=v^2 n$, with $v$ the expectation value of the
Higgs field and $n>0$, is proportional to the soliton mass and
$M_{Pl}$ is the three-dimensional Planck mass. 
The geometry has deficit angle $\delta=2\pi M/M_{Pl}$, 
and the soliton saturates the BPS bound.
The gravitino gives rise to an
Aharanov-Bohm phase that exactly cancels the geometric phase associated
to the deficit angle of the conical singularity.  However, the fermionic
zero mode is not normalizable and is absent from the physical spectrum.
Thus, there is no $N=1$ supermultiplet of the unbroken supersymmetry.
In this way, Witten's observation holds and the bose-fermi degeneracy of
the excited states is lifted even though the solitonic ground state
has zero vacuum energy.

\section{Deconstructing Gauge Theories: A Summary}

Before we discuss the case of $(3+1)$-dimensional gravity, let us review
the gauge theory case from a slightly different point of view than the
original presentation \cite{nima}.

Consider a gauge theory action
\beq
S=-\frac{1}{2g^2_{d}}\int d^{d-1}x dy\ \tr F_{AB}^2(x,y).
\eeq
We use the notation $x^A\equiv \{ x^\mu,y\}$. We wish to arrive at a theory on 
the
space  ($\bR^{d-1}\times\Gamma$). 
We then latticize $y$ with lattice spacing $a$:
\beq
S=-\frac{a}{2g_d^2}\int d^{d-1}x\sum_{j} \tr \left( F_{\mu\nu,j}^2(x)+2F_{5\nu,j}^2\right).
\eeq
In the continuum,
\beq
F_{5\nu,j}=\partial_5 A_\nu-\partial_\nu A_{5}+i[A_{5},A_\nu].
\eeq
We define a link variable in the usual way:
\beq
U_{j,j+1}=\exp\left( i\int dy A_5(x,y)\right)\simeq 1+iaA_{5,j}+\ldots.
\eeq
Therefore
\beq
F_{5\nu,j}\simeq \frac{1}{a}\left(A_{\nu,j+1}-A_{\nu,j}\right)+\frac{i}{a}\partial_\nu U_{j,j+1}
+\frac{1}{a} (U_{j,j+1}-1)A_{\nu,j+1}-\frac{1}{a}A_{\nu,j} (U_{j,j+1}-1)+\ldots
\eeq
(the ellipses contains less relevant terms for small $a$) and so
\beqn
F_{5\nu,j}&\simeq& \frac{1}{a}\left(
i\partial_\nu U_{j,j+1}
+ U_{j,j+1}A_{\nu,j+1}-A_{\nu,j} U_{j,j+1}
\right)\\
&=& \frac{i}{a}D_\nu U_{j,j+1}.
\eeqn
As a result:
\beqn
S&=&-\frac{a}{2g_d^2}\int d^{d-1}x\sum_{j} \tr  \left(F_{\mu\nu,j}^2(x)-\frac{2}{a^2}|D_\nu U_{j,j+1}|^2\right)\\
&=&-\frac{1}{2g_{d-1}^2}\int d^{d-1}x\sum_{j} \tr  F_{\mu\nu,j}^2(x)
+f_\pi^2\int d^{d-1}x\sum_{j} \tr |D_\nu U_{j,j+1}|^2,
\eeqn
where
\beq
\frac{1}{g_{d-1}^2}=\frac{a}{g_d^2},\ \ \ \ \
f_\pi^2=\frac{1}{ag_d^2}=\frac{1}{a^2g_{d-1}^2}.
\eeq

Let us redo this computation, as there is actually some trickery involved
in the above continuum calculation.

To achieve this, we put the entire theory on 
a lattice and take the
continuum limit in all but the $y$ direction (which retains 
lattice spacing $a$):
\beql{lattact}
S_{latt}=\sum_P \scale_P\left( -1+\frac{1}{2N}\tr (U_P+U_P^\dagger)\right),
\eeq
where $\tr\,1=N$ and $\scale$ is an appropriate numerical scaling factor.  $P$ denotes a plaquette, which we can think of as a sum
over lattice points, and a sum over pairs of directions $\hat A,\hat B$ and
\beq 
U_{AB}(n)=U_A(n)U_B(n+\hat A)U^\dagger_A(n+\hat B)U^\dagger_ B(n),
\eeq
where $U_A(n)$ is a link field, which in the continuum limit goes to the
Wilson line. In the present case, we split the index $A$ into $\mu,5$ with 
lattice
spacings $\epsilon,a$. There are two types of terms in eq.\ \eq{lattact}:
\beql{lattacta} 
S_{latt}=\sum_{\mu\nu} \scale_{\mu\nu} \left( -1+\frac{1}{2N}\tr 
(U_{\mu\nu}+U_{\mu\nu}^\dagger)\right)+\sum_{\mu} \scale_{\mu 5}\left( 
-1+\frac{1}{2N}\tr (U_{\mu 5}+U_{\mu 5}^\dagger)\right).
\eeq
The first term will go in the continuum limit to
$-\frac{\scale_{\mu\nu}}{2}\epsilon^4 \tr F_{\mu\nu}^2$, while the second term
yields $-\frac{\scale_{\mu 5}}{2}\epsilon^2\tr D_\mu U_5 (D_\mu U_5)^\dagger$,
where
\beq
D_\mu U_5=\partial_\mu U_5+i(A_{\mu,j}U_5-U_5 A_{\mu,j+1}).
\eeq
Thus, we arrive at
\beq
S_{latt}= - \sum_{\mu\nu}\frac{\scale_{\mu\nu}}{2}\epsilon^{5-d}\int d^{d-1} x \sum_j \left\{ \tr 
F_{\mu\nu,j}^2(x)+\frac{\scale_{\mu 5}}{\scale_{\mu\nu}\epsilon^{2}} \tr |D_\mu 
U_5|^2\right\}.
\eeq
By appropriate scalings of the parameters, we may obtain:
\beqn
S &=&-\frac{1}{2g_{d-1}^2}\int d^{d-1}x\sum_{j} \tr  F_{\mu\nu,j}^2(x) 
+f_\pi^2\int d^{d-1}x\sum_{j} \tr |D_\nu U_5|^2
\eeqn
with $f_\pi=1/(g_{d-1}a)$.

We note that the link field $U_5$ is a bifundamental, transforming as 
$U_5\to V_j U_5 V_{j+1}^{-1}$.
The essential non-perturbative information used at this point is that
fermion condensation can induce the effective $\sigma$-model action in the IR.
This then points to the degrees of freedom of an $SU(N)$ quiver theory
\cite{nima}.  Thus the UV completion of a non-renormalizable five-dimensional
gauge theory is a very specific quiver theory \cite{nima}.

Unfortunately, if we wish to obtain the continuum limit in the quantum theory, 
we must take $a\to0$ holding $g_d$ fixed. This scales $g_{d-1}\to\infty$, and 
thus the infrared dynamics is in fact significantly different than the 
classical theory would indicate.

\section{Towards a UV Completion of $(3+1)$-dimensional Einstein-Hilbert 
Gravity}

Now we are ready to address the question of deconstruction of
$(3+1)$-dimensional gravity from the point of view of Ref.\ \cite{nima}.
As we have reviewed above, the scenario has been
initially applied to certain non-renormalizable gauge theories.  Given a
certain set of similarities between pure gravity and
non-abelian gauge theories it is natural to wonder whether the deconstruction
techniques can be successfully applied to gravity \cite{others}.
Along the lines of Ref.\ \cite{nima}, in this section we shall construct a
lattice of coupled $(2+1)$-dimensional theories, which in the IR exhibits
the features of $3+1$ gravity.

There exist many similarities between gravity and gauge theory.  These are
evident, for example, in the MacDowell-Mansouri approach \cite{cmm}; the
approach to $3+1$ gravity based on Ashtekar variables \cite{loopqg}; the
approach to $(2+1)$-dimensional general relativity based on 
CS theory \cite{wittencs}; the close relation between the topological
BF theory and gravity in any dimension \cite{bf}; the appearance of an
induced Chern Simons theory in the context of $(3+1)$-dimensional gravity on manifolds with
a boundary \cite{lee}, etc.

Given the fact that deconstruction provides a procedure for defining UV
completions of certain, in principle, non-renormalizable field theories,
it is only natural to ask whether similar reasoning can be applied to
$(3+1)$-dimensional gravity, while remembering that the $(2+1)$-dimensional
(pure gravity) theory is well-defined.  In other words, is it possible to
deconstruct $(2+1)$-dimensional CS coupled to certain matter fields into
a pure $(3+1)$-dimensional gravity?

Many things point to the possibility that four-dimensional gravity can
be defined in terms of purely three-dimensional data.
For example, three-dimensional CS actions appear as natural boundary terms in the
connection formulation of four-dimensional theory \cite{lee}, as well as
in the relation between the BF topological theory and $(3+1)$-dimensional
general relativity.\footnote{Note also,
that in the framework of the $AdS$/CFT correspondence four-dimensional
Poincar\'e supergravity data can be reconstructed from three-dimensional
conformal supergravity data \cite{joachim}.}

There even exists a theorem concerning dimensional reduction
in classical general relativity which states that for the case
of space-like Killing fields $3+1$ gravity can be rewritten as $2+1$ gravity
coupled to a non-linear $SO(2,1)$ $\sigma$-model \cite{grbook, aashtekar}.
More precisely, in a classical background with a space-like isometry, the metric can be
put in the form
\beq
ds^2=N^2(x)dr^2+\hat g_{ab}(x)(dx^a+N^a(x)dr)(dx^b+N^b(x)dr).
\label{eq:metric}
\eeq
For the case of $3+1$ dimensions, the vacuum classical equations of
motion are particularly simple \cite{grbook, aashtekar}, and reduce to
$2+1$-dimensional gravity coupled to scalar fields. Let us disregard the shift 
fields for the sake of simplicity. The equations of motion
can be written in the following form
\beq
{\hat R}^{(3)}_{ab}(x) = 2 {\hat\nabla}_a\phi(x)\cdot {\hat\nabla}_b \phi(x)
,\ \ \ \ \ \ \ {\hat g}^{ab}{\hat\nabla}_a {\hat\nabla}_b \phi =0.
\eeq
Here ${\hat R}^{(3)}_{ab}$ and ${\hat\nabla}_a$ are the Ricci tensor and
the covariant derivative associated with ${{\hat g}}_{ab}$, and $\phi$ is a 
scalar field arising after a field redefinition of $N$ and $\hat g$ \cite{aashtekar}.
By suitable rescalings, we can bring this to the form\beq
\hat R^{(3)}_{ab} - \frac{1}{2}\hat g_{ab} \hat R^{(3)} = 8\pi G_3\ T_{ab},
\eeq
where the covariantly conserved energy momentum tensor reads
\beq
T_{ab} = {\hat\nabla}_a \Phi {\hat\nabla}_b \Phi - 
\frac{1}{2}\hat g_{ab} {\hat\nabla}_c \Phi {\hat\nabla}^c \Phi.
\eeq
Therefore, the $(3+1)$-dimensional vacuum equations in the presence of a
space-like Killing field are equivalent to the $(2+1)$-dimensional gravity
coupled to a scalar field, illustrating the more general theorem
stated in Ref.\ \cite{grbook}.

This is of course only an on-shell observation.  We claim that in the quantum theory,
a similar condition holds {\em locally} and applies at the level of the
action and the path integral.

To argue this, we start from the classical formulation of gravity using the $d$-bien and
spin connection as variables, with action
\beq
S_{EH}=\frac{1}{G_d}\int d^dx\ \epsilon_{a_1\ldots a_d}\epsilon^{A_1\ldots 
A_d}
e_{A_1}^{a_1}\ldots e_{A_{d-2}}^{a_{d-2}}{R_{A_{d-1}A_d}}^{a_{d-1}a_d},
\eeq
where (note $G_d$ has units of $m^{2-d}$, as $[{\omega^a}_b]=[{R^a}_b]=1, 
[e^a]=L$ as forms)
\beq 
R=d\omega+\omega\wedge\omega.
\eeq
We are  careful to distinguish various indices: we are on a manifold $M$,
with tangent bundle $TM$. The indices $A,B,...$ label vectors in $TM$. We
also have a vector bundle $V$ with structure group $SO(d-1,1)$, which we
will assume is (more or less) isomorphic to $TM$. Indices for vectors in
$V$ will be given by $a,b,...$. These latter indices eventually will be
thought of as ``gauge'' indices.

Let us focus on $d=4$. We then have
\beq
S_{EH}=\frac{1}{G_4}\int \epsilon_{abcd}e^a\wedge e^b\wedge R^{cd}=
\frac{2}{G_4}\int d^4x\epsilon^{\mu\nu\lambda}\epsilon_{abcd}\left(e_3^a 
e_\mu^b {R_{\nu\lambda}}^{cd}-e_\mu^a e_\nu^b {R_{\lambda 3}}^{cd}\right).
\eeq
To go to the lattice, we have many options. In the gauge theory case,
gauge covariance was maintained throughout, and the lower dimensional
theory had gauge group $G^N$. Analogously, the simplest lattice action to
take in the case of gravity would be to keep $SO(3,1)$ invariance. We can
regard $\omega$ as an $SO(3,1)$ connection and replace it by plaquettes in
the lattice version. To begin,\footnote{It seems also that there could be a 
formalism where
we treat $e,\omega$ as connections for $ISO(3,1)$, and thus introduce link 
fields corresponding to $e_3$ as well. We will not follow that approach here.}
 we will suppose that the vierbein remains
as a site field. The appropriate thing to do then is replace $S_{EH}$
by a lattice version in which the curvature $R_{AB}$ is replaced by 
$Im\ U_{AB}$. Along the lines of the calculations in the Yang-Mills
theory, we find:
\beqn
Im\ U_{\mu 3} &=& \epsilon J_\mu+\ldots,\\
Im\ U_{\mu\nu} & =& -2\epsilon^2 R_{\mu\nu}+\ldots
\eeqn
where 
\beq
J_\mu(x,j)=i\left( D_\mu U_3 \cdot U_3^\dagger-U_3 (D_\mu U_3)^\dagger\right).
\eeq
Taking $\epsilon\to0$, we will obtain
\beq
S_{EH}=\frac{2a}{G_4}\int d^3x\ \sum_j 
\epsilon_{abcd}\epsilon^{\mu\nu\lambda}\left(
e_3^a e_\mu^b {R_{\nu\lambda}}^{cd}+\scale e_\mu^ae_\nu^b 
{J_{\lambda}}^{cd}\right).
\eeq
Note that in writing this action, we have essentially forced $SO(3,1)_{(j)}$
invariance at each site $j$. Although $U_3$ is a link field, and thus
transforms as $U_3\to \Lambda_j U_3\Lambda_{j+1}^{-1}$, the current is a
tensor only under the local slice, $SO(3,1)_{(j)}$.

Thus, we have an action of the form
\beq
S=\frac{1}{G_3}\sum_j\int_{{\mathcal M}_j} \epsilon_{abcd}\left[ \varphi^a 
e^b\wedge R^{cd}+f e^a\wedge e^b\wedge J^{cd}\right]
\eeq
where we have dropped the index $j$ on fields and written $\varphi\equiv
e_3$ and $f\equiv 1/a$. This action manifestly possesses ${\rm Diff}_3\times 
SO(3,1)$
invariance.  We can introduce a four-dimensional cosmological constant
as well:
\beq
S=\frac{1}{G_3}\sum_j\int_{{\mathcal M}_j} \epsilon_{abcd}\left[ \varphi^a 
e^b\wedge R^{cd}+\lambda \varphi^a e^b\wedge e^c\wedge e^d+f e^a\wedge 
e^b\wedge J^{cd}\right].
\label{eq:slat}
\eeq
This looks like a $(2+1)$-dimensional ``gauge theory'' coupled to a
current $J$.  Note, however, the Latin indices are $(3+1)$-dimensional, and 
thus this is not in any sense ``$2+1$ gravity''. Furthermore, there are $N$
copies of the symmetry group.

The UV theory could also possess $\sigma$-model terms such as
\beql{curvsq}
S_\sigma=\sum_j\int_{{\mathcal M}_j} \epsilon_{abcd}\left[ J^{ac}\wedge 
*J^{bd}\right]
\eeq
as well as other higher order terms. Our point of view here is that in the
UV, we can treat the theory as containing just a set of currents with kinetic terms if
necessary. As we go to the
IR (the continuum limit), the current kinetic terms become {\it irrelevant}
({\em e.g.}, eq.\ \eq{curvsq} becomes a curvature-squared term), leaving
only the Einstein-Hilbert action.

Of course, an important aspect of this is that the continuum limit must exist in 
some sense.
In fact, the original four-dimensional action is an effective theory, which is 
certainly only valid for probes at length scales\footnote{We use the notation 
$L_d$ for the $d$-dimensional Planck length.} $L>>L_4$. Thus, if $a<L_4$, the 
available probes cannot tell the difference between the lattice theory and the 
continuum.  Consequently, the region of strong three-dimensional coupling can be 
avoided, while staying within the region of validity of the four-dimensional 
theory. Essentially, classically the three and four-dimensional theories are 
equivalent, as constructed. We propose that this remains true even in 
the quantum theory. Furthermore, we will have to suppose that the value of the 
four-dimensional cosmological constant is given by its limiting
three-dimensional value. This seems obvious if we don't have to strictly take
the continuum limit.

We could now try to proceed further and reduce $SO(3,1)$ to $SO(2,1)$, to
make the theory look gravitational in $2+1$ as well. We start by just
segregating indices:
\beqn
S=\sum_j\int_{{\mathcal M}_j} \epsilon_{\alpha\beta\gamma}\left[  (\varphi^3 
e^\alpha-\varphi^\alpha A)\wedge (R^{\beta\gamma}-\omega^{\beta}\wedge 
\omega^{\gamma})-2\varphi^\alpha e^\beta\wedge 
(D\omega)^{\gamma}\right.\nonumber \\ \left.+\lambda \varphi^3 e^\alpha\wedge 
e^\beta\wedge e^\gamma-3\lambda\varphi^\alpha A\wedge e^\beta\wedge 
e^\gamma-2f_\pi e^\alpha\wedge e^\beta\wedge J^{\gamma}+2f_\pi A\wedge 
e^\alpha\wedge J^{\beta\gamma}\right],
\eeqn
where $A\equiv e^3$, $(D\omega)^{\gamma}\equiv 
d\omega^\gamma+{\omega^\gamma}_\delta\wedge\omega^{\delta}$, 
$\omega^\alpha\equiv\omega^{\alpha,3}$,
and $J^\gamma\equiv J^{\gamma,3}$. 
Note that with the assumed lattice,  there are natural vevs:
\beq
\langle \varphi^3\rangle = 1,\ \ \ \ \ 
\langle \varphi^\alpha\rangle = 0,\ \ \ \ \ \langle A_\mu\rangle=0,
\eeq
which put the background metric in the form appropriate to the chosen lattice
\beq
ds^2=a^2(\Delta j)^2+ds_{2,1}^2 (j).
\eeq
This metric is just a discretized form of the canonical ``ADM''
metric \cite{MTW}
\beq
ds^2= N^2(x,r)dr^2+g_{\mu\nu}(x,r) [dx^\mu +N^{\mu}(x,r)dr][dx^\nu 
+N^{\nu}(x,r) dr].
\eeq
Here $r$ denotes the continuum limit of the discretized lattice direction.
In this discretized form the shift vector has been expanded
around zero.
In writing down eq.\ \eq{slat}, the lattice action for $3+1$ gravity, we
have set the shift vector to zero.  Locally, we can always do this, but,
generically, we cannot turn this into a global choice.

Thus it is natural to expand in fluctuations around this vev, (fluctuations in 
$\varphi^a, A_\mu$ correspond to modifications in the
shape of the lattice) and we obtain:
\beq S=S_{\oplus EH}+S_{int}
\eeq
where 
\beq
S_{\oplus EH}=\frac{1}{G_3}\sum_j\int_{{\mathcal M}_j} 
\epsilon_{\alpha\beta\gamma}\left[   e^\alpha\wedge R^{\beta\gamma}+\lambda 
e^\alpha\wedge e^\beta\wedge e^\gamma\right]
\eeq
and
\beqn
S_{int}=
\sum_j\int_{{\mathcal M}_j} \epsilon_{\alpha\beta\gamma}\left[  
-e^\alpha\wedge \omega^{\beta}\wedge \omega^{\gamma}+(\varphi^3 
e^\alpha-\varphi^\alpha A)\wedge (R^{\beta\gamma}-\omega^{\beta}\wedge 
\omega^{\gamma})-2\varphi^\alpha e^\beta\wedge 
(D\omega)^{\gamma}\right.\nonumber \\ \left.+\lambda \varphi^3 e^\alpha\wedge 
e^\beta\wedge e^\gamma-3\lambda\varphi^\alpha A\wedge e^\beta\wedge 
e^\gamma-2f e^\alpha\wedge e^\beta\wedge J^{\gamma}+2f A\wedge e^\alpha\wedge 
J^{\beta\gamma}\right].
\eeqn
In addition, we would add matter fields to $S_{int}$.

Provided the $(2+1)$-dimensional
currents $J^{\mu \nu}$ can be dynamically induced via some non-perturbative
mechanism from some other well-defined degrees of freedom,
in the deep UV one would be left only with ($N$ copies) of the 
$(2+1)$-dimensional CS
term coupled to these $(2+1)$-dimensional degrees of freedom.  
In the intermediate range of scale we get $N$ copies
of linked $(2+1)$-dimensional CS theories coupled to $(2+1)$-dimensional
currents.\footnote{Our concluding picture resembles somewhat that of
Ref.\ \cite{crane}.} In the IR we recover the full $(3+1)$-dimensional
general relativity.\footnote
{One could also entertain the possibility of simply starting with $2+1$ gravity coupled 
to appropriate fields. In this case, $SO(3,1)$ would have to be an accidental symmetry 
of the IR.}
It should be pointed out that the recovery of the full diffeomorphism group in $3+1$ 
dimensions from this construction is rather non-trivial given that we work on the 
lattice and because the IR physics lies in the strong coupling regime.
In the very deep IR, {\em i.e.}\ when $N$ is finite and the wavelength 
exceeds the lattice size, the physics is, of course, again $(2+1)$-dimensional.

Notice that we have the right number of degrees of freedom needed to
reproduce the $(3+1)$-dimensional theory.  These degrees of freedom come
from the matter fields coupled to the CS theory.  Thus our formulation does
provide a quantum mechanical version of the classical theorems discussed
above \cite{grbook, aashtekar}.

Given that the theory is $(2+1)$-dimensional in the UV, one might wonder
whether the picture is compatible with the Bekenstein-Hawking bounds on
entropy \cite{bhent}.  Let us suppose that the $(2+1)$-dimensional matter
fields are local.  The coupling of $2+1$ gravity to matter is of the 
general form
\begin{equation}
S_{EH} = \frac{1}{G_3} \int d^3x\, \sqrt{-g^{(3)}}\, (R^{(3)} + {\cal L}_{\rm matter}).
\end{equation}
The entropy of local matter degrees of freedom scales as the two-dimensional 
area.  As there are $N$ copies, we have
\begin{equation}
{\cal S} \propto \frac{NA}{G_3}.
\end{equation}
This expression does not have the correct mass dimension.  The usual
prescription for dimensional reduction tells us that the pre-factor should
be $1/G_3 L$, where $L=Na$ is the size of the fourth (lattice) dimension.
Thus, on heuristic grounds,
\begin{equation}
{\cal S} \simeq \frac{NA}{G_3 L} = \frac{A}{G_3 a} = \frac{A}{G_4}, 
\end{equation}
which reproduces the Bekenstein-Hawking scaling in $3+1$ dimensions.
Of course, dimensional analysis does not reproduce the numerical factor
of $1/4$ in the entropy formula.

As we will argue in the next concluding section, the above observations
are enough to argue that Witten's mechanism for vanishing of the
$(2+1)$-dimensional cosmological constant can be lifted to $3+1$ dimensions.

\section{The Vanishing Cosmological Constant Deconstructed}

Now we argue that Witten's argument for the vanishing of $(2+1)$-dimensional
vacuum energy can be deconstructed as follows:

1) Assume a local spatial foliation of $(3+1)$-dimensional spacetime.

2) Deconstruct the vacuum part of pure $(3+1)$-dimensional gravity from ($N$ copies of)
$(2+1)$-dimensional general relativity coupled to certain $(2+1)$-dimensional
matter fields represented in terms of currents as in the preceding section.
Assume that $3+1$-dimensional sources can be defined in terms of a 
deconstructed $2+1$-dimensional theory. For sources represented
by gauge fields this should be possible given the discussion\footnote{One might ask why a $(3+1)$-dimensional
theory with a well
defined $(3+1)$-dimensional UV behavior, such as the Standard Model, should be
defined in terms of $(2+1)$-dimensional data. The point here is
that both the deconstructed $(2+1)$-dimensional and the intrinsic
$(3+1)$-dimensional UV definitions lead to the same IR physics,
and as such are indistinguishable at long distances.} of 
section 3.

3) In the deep UV we have ($N$ copies of) $2+1$ gravity coupled to some
$(2+1)$-dimensional sources.  Whatever the matter content of this
$(2+1)$-dimensional theory is, we know that the resulting geometry has to
be conical.  Thus Witten's argument applies: the vacuum is supersymmetric,
yet the excited states are not.

4) In the range of intermediate scales, we have $N$ linked copies
of $2+1$ gravity coupled
to $(2+1)$-dimensional currents.
Once again, the resulting $(2+1)$-dimensional geometry is conical. Thus Witten's argument holds in
the region between UV and IR.

Finally notice that on dimensional grounds, 
the mass splitting should be 
inversely proportional to the three-dimensional Newton constant and 
should vanish at zero deficit angle.  
We take $\Delta m \simeq \delta/G_N$.
Thus as long as the three-dimensional Newton constant
is of order one as the continuum limit is taken,
and the deficit angle (on each local three-dimensional
slice) is taken to scale as the inverse of the lattice spacing,
the fermi-bose splitting will be finite in the infrared. These remarks may be tested by
examination of the example of Ref. \cite{bbs}.\footnote{For example \cite{bbs}, the deficit angle produced by a mass $M$
is $\delta = 2\pi ML_3$.  Thus, the mass difference (at one-loop) between
fermions and bosons should be proportional 
to $g^2 \delta/G_N = 2\pi g^2M$, where $g$ is the interaction strength. 
In a realistic model the mass $M$ should be deconstructed to be
of the order of a TeV.} 

According to the outlined argument the vacuum energy is zero in the UV,
and also some place in between UV and IR.  But does it remain zero in the IR?
That is difficult to say, given the fact
that the three-dimensional coupling has
to be of order one, but the physical picture would be that as one takes the
lattice spacing to zero, one still has in principle an infinite number
of  $(2+1)$-dimensional matter theories strongly coupled to $2+1$
gravity.

Essentially we have a deconstruction
of $(2+1)$-dimensional conical singularities to one-dimensional,
string-like singularities in every local patch of $3+1$ dimensions.
Thus we again end up with a claim that the vacuum state can be made
supersymmetric and yet the excited states do not fall into supermultiplets
because of the non-existence of the global supercharge due to the presence
of the string-like defects which create a deconstructed version of the
asymptotically conical three-dimensional geometry.\footnote{As reviewed in
Ref.\ \cite{aashtekar}, the information about the mass resides in $(2+1)$-dimensions
in the ``zeroth order'' behavior of the metric at infinity.
This should be contrasted to the situation in $(3+1)$-dimensions 
where the analogous information about the mass resides
in the leading $1/r$ deviations from the Minkowski metric near 
infinity \cite{aashtekar}.}

Within this framework the four-dimensional cosmological constant is essentially
determined by the value of the three-dimensional cosmological constant.
In the supersymmetric scenario the latter
is zero, and so is the four-dimensional one. Yet the excited states are
non-supersymmetric due to the non-existence of a global supercharge.

Our actual calculations in this paper have all been non-supersymmetric. They
may easily be generalized however --- for example, the MacDowell-Mansouri
approach \cite{cmm} provides a unified geometric formulation of supersymmetry
and gravity with the curvature constructed from the spin connection,
the vierbein, and the gravitino. The analysis presented in section 4
applies also in this situation.  It would be very interesting to study
the deconstruction of this theory explicitly.

We conclude this article with an obvious question: assuming that
the ultraviolet completion of $(3+1)$-dimensional gravity
is indeed given in terms of $(2+1)$-dimensional gravity
coupled to $(2+1)$-dimensional matter as we have argued above, what
are the most immediate observational consequences
and constraints, in the sense of 
$(3+1)$-dimensional gravity being 
modified at very short distances?

\vspace{0.5in}

\section*{Acknowledgments}
\noindent We thank A.\ Ashtekar, C.\ Bachas, V.\ Balasubramanian, J.\
de Boer, R.\ Emparan, W.\ Fischler, L.\ Freidel, E.\ Gimon, N.\ Kaloper,
P.\ Kraus, D.\ Marolf, R.\ Myers, A.\ Naqvi, R.\ Rattazzi, L.\ Smolin,
and J.~P.\ van der Schaar for conversations and comments.
VJ thanks the High Energy Group at the University of Pennsylvania for
their generous hospitality.  DM thanks the Perimeter Institute and the
Lorentz Center of Leiden University for their hospitality.
The work of RGL is supported in part by the U.S.\ Department of Energy
under contract DE-FG02-91ER40677.  The work of VJ and DM is supported in
part by the U.S.\ Department of Energy under contract DE-FG05-92ER40709.

\end{document}